\documentclass{article}
\usepackage[utf8]{inputenc}
\usepackage{graphicx, subcaption}
\usepackage{enumerate}
\usepackage{amsmath, mathtools, amssymb, amsthm}
\usepackage[margin=1in]{geometry}
\usepackage{soul}
\usepackage{bm}
\usepackage{authblk}
\usepackage{multirow}

\usepackage{listings}
\usepackage{xcolor}

\definecolor{codegreen}{rgb}{0,0.6,0}
\definecolor{codegray}{rgb}{0.5,0.5,0.5}
\definecolor{codepurple}{rgb}{0.58,0,0.82}
\definecolor{backcolour}{rgb}{0.95,0.95,0.92}

\lstdefinestyle{mystyle}{
    backgroundcolor=\color{backcolour},   
    commentstyle=\color{codegreen},
    keywordstyle=\color{magenta},
    numberstyle=\tiny\color{codegray},
    stringstyle=\color{codepurple},
    basicstyle=\ttfamily\footnotesize,
    breakatwhitespace=false,         
    breaklines=true,                 
    captionpos=b,                    
    keepspaces=true,                 
    numbers=left,                    
    numbersep=5pt,                  
    showspaces=false,                
    showstringspaces=false,
    showtabs=false,                  
    tabsize=2
}

\newcommand{\myindent}{\hspace*{0.4 cm}}

\renewcommand{\bf}[1]{\mathbf{#1}}
\newcommand{\X}[1]{ \big[X_{#1} \big]}

\date{\vspace{-5ex}}

\title{Sparse Data Structures for Efficient State-to-State Kinetic Simulations}
\author[1]{Ayoub Gouasmi}
\author[2]{Scott M. Murman}
\affil[1]{Oak Ridge Associated Universities}
\affil[2]{NASA Ames Research Center, NASA Advanced Supercomputing Division}

\begin{document}
\maketitle

\begin{abstract}
Higher-fidelity entry simulations can be enabled by integrating finer thermo-chemistry models into compressible flow physics. One such class of models are State-to-State (StS) kinetics, which explicitly track species populations among quantum energy levels. StS models can represent thermo-chemical non-equilibrium effects that are hardly captured by standard multi-temperature models. However, the associated increase in computational cost is dramatic. For implicit solution techniques that rely on standard block-sparse representations of the Jacobian, both the spatial complexity and the temporal complexity grow quadratically with respect to the number of quantum levels represented. We introduce a more efficient way to represent the Jacobian arising in first-order implicit simulations for compressible flow physics coupled with StS models. The key idea is to recognize that the density of local blocks of the Jacobian comes from rank-one updates that can be managed separately. This leads to a new Jacobian structure, consisting of a fully-sparse matrix and block-wise rank-one updates, whose overall complexity grows linearly with the number of quantum levels. This structure also brings forth a potentially faster variation of the block-Jacobi preconditioning algorithm by leveraging the Sherman-Morrison-Woodbury inversion formula.  
\end{abstract}

\section{Introduction}
\myindent For several decades, Computational Fluid Dynamics (CFD) has been a cornerstone technology in space missions. Its importance stems from the inherent limitations of ground-based testing, which cannot fully replicate the extreme conditions of space entry, and their considerable cost. Through the use of CFD simulations, research scientists and engineers have been able to achieve more efficient and robust designs for Entry, Descent, and Landing (EDL) technology. One notable application has been the accurate prediction of heating loads during entry, which is crucial for the development of effective heat shields \cite{Candler_review, Park_book, Candler_shock, Park_2T, Gnoffo1989} . Yet, CFD still remain an active field of research with a plethora of challenges in physical modeling, numerical algorithms and high performance infrastructure \cite{CFD_vision_2014, CFD_vision_2021}.  \\
\myindent We are interested in some of the challenges that arise with the use of more refined thermo-chemistry models in the CFD simulation of entry physics \cite{Park_Air, Park_Mars}. Without chemistry, entry calculations tend to grossly over-predict heating loads and misplace shock structures. With chemistry, CFD calculations are more accurate but the computational workload needed to make more reliable predictions increases dramatically. This is in large part due to the introduction of temporal chemical scales that are much smaller than those associated with compressible flow phenomena. The stiffness of the discrete system to evolve increases to a point where implicit temporal schemes are preferred over explicit ones \cite{Curtiss1952, Park1985}. \\
\myindent A simple and fairly representative case problem to investigate thermo-chemical non-equilibrium effects in entry physics is the steady-state calculation of hypersonic flow past a blunt-body in two dimensions. As a reference solution approach, we consider an implicit steady-state scheme, where a first-order Backward Euler scheme is used to advance a pseudo-transient initial solution until the norm of a first-order finite-volume residual is below a threshold value. Each pseudo time step involves the solution of a fully-discrete nonlinear system of equations. For this a standard Newton-GMRES method is used, and a large portion of the compute time is spent solving linear system determined by the Jacobian. The present work is exclusively concerned with the data structures used to explicitly represent the full Jacobian. A standard representation uses block-sparse matrices where only non-zero Jacobian data blocks are stored in memory, along with index information. \\
\myindent Standard thermo-chemistry models typically represent the gas mixture as thermally perfect, with the specific internal energy of each species being a function (polynomial or analytic) of one single temperature, and represent chemical reactions using mass action models where the rate functions depend solely on temperature and are related through chemical equilibrium constants. From the standpoint of statistical thermodynamics, this assumes that the internal distribution of the gas particles among quantum energy levels follows an equilibrium Boltzmann distribution \cite{Vincenti}. Dynamically, this means that the non-equilibrium quantum distributions that arise immediately behind the shock equilibrate much faster than chemical processes \cite{Anderson}. Under low density and high temperature conditions characteristic of entry missions, this does not hold and both thermal and chemical processes will not be predicted accurately \cite{Candler_shock, Candler_review}. State-to-State (StS) kinetic models \cite{Jaffe2010, Armenise1996, Josyula2015, Josyula2017, Bonelli2017, Ninni2022, CapitelliBook} explicitly tracking quantum populations provide a higher level of fidelity, but their added cost is considerable. Integrating StS kinetics models dramatically increases the number of governing equations. Both the temporal complexity and the spatial complexity of discrete Jacobian operations will grow quadratically with respect to the number of quantum levels tracked. \\
\myindent In this work, we introduce more efficient data structures to store and operate on Jacobian matrices arising in implicit simulations of coupled inviscid fluid and StS kinetics processes. The key idea is to recognize that the density of local blocks of the Jacobian comes from rank-one updates that can be managed separately. This leads to a new Jacobian structure, consisting of a fully-sparse matrix and block-wise rank-one updates, whose overall complexity grows linearly with the number of quantum levels.  \\
\myindent The paper is organized as follows. In section 2, we introduce the compressible flow model and the thermo-chemistry model. In section 3, we present the reference discretization for a steady-state implicit first-order finite-volume solution of the governing equations, and we discuss the associated Jacobian structure. In section 4, we show how to re-organize Jacobian data from discretization components using rank-one updates. We do so locally first, looking at the respective contributions from the inviscid flux and from the chemical model to individual Jacobian blocks. From there, we then outline the resulting Jacobian representation as a block r1-sparse matrix. In section 5, we discuss a working implementation of our Jacobian data structure and demonstrate and demonstrate its performance as the number of quantum levels grows. Finally, in section 6, we suggest a variation of the standard block-Jacobi preconditioning algorithm that naturally arises when combining our novel Jacobian representation with the Sherman-Morrison-Woodbury matrix inversion formula.

\section{Physical Model}
\subsection{Compressible Flow Model}
\myindent We consider a multi-component reacting compressible flow system involving $N_{s}$ species in two spatial dimensions. This can be modeled using a system of $m$ Partial Differential Equations (PDE) \cite{Giovangigli}:
\begin{equation}\label{eq:PDE_base}
       \frac{\partial \bf{u}}{\partial t} \ + \ \frac{ \partial \bf{f}_{x}}{\partial x} \ + \ \frac{ \partial \bf{f}_{y}}{\partial y} \ = \ \bf{\Omega}.
\end{equation}
The governing equations describe the conservation of species mass, momentum and total energy. The number of equations $m$ is related to the number of species by $m = N_{s} + 3$. The temporal flux $\mathbf{u}$ and spatial fluxes $\big( \bf{f}_{x}$, $\bf{f}_{y}\big)$ are given by:
\begin{align*}
    \bf{u} \ :=& \ \begin{bmatrix} \big( \rho_k \big)_{1 \leq k \leq N_{s}} & \rho u & \rho v & \rho e^{tot} \end{bmatrix}^T, \\
    \bf{f}_{x} \ :=& \ \begin{bmatrix}  \big( \rho_k u \big)_{1 \leq k \leq N_{s}}  & \rho u^2 + p & \rho u v &(\rho e^{tot} + p)u \end{bmatrix}^T, \\
    \bf{f}_{y} \ :=& \ \begin{bmatrix}  \big( \rho_k v \big)_{1 \leq k \leq N_{s}} & \rho u v & \rho v^2 + p & (\rho e^{tot} + p)v \end{bmatrix}^T, 
\end{align*}
where subscript $k$ refers to the species index. $\rho_k$ denotes the partial density. $\rho := \sum_{k=1}^N\rho_k$ denotes the total density. The flow velocity in the x and y directions are given by $u$ and $v$, respectively. The pressure $p$ is given by the ideal gas law:
\begin{equation*}
p := \sum_{k=1}^N\rho_k  r_k T, \ r_k = \frac{R}{M_k},
\end{equation*}
where $M_k$ is the species molar mass, $R$ is the gas constant and $T$ is the temperature. \\
\myindent $e^{tot}$ denotes the specific total energy of the gas mixture. It is defined by the relation:
\begin{equation*}
    e^{tot} \ := \ e \ + \ \frac{1}{2} \big( u^2 \ + \ v^2),
\end{equation*}
where $e$ denotes the specific internal energy of the gas mixture, which is typically represented as a linear combination of the species individual species energies $\big(e_{k}\big)_{1 \leq k \leq N_{s}}$.
\begin{equation*}
\rho e = \sum_{k=1}^{N_{s}} \rho_k e_k, \ e_k = e_k(T).
\end{equation*}
For calorically perfect gases, the specific internal energies are linear functions of $T$. For thermally perfect gases, the dependency with respect to temperature becomes nonlinear, with either curve-fit models or analytical expressions available for use in the literature \cite{Chalot1990, Gnoffo1989}. Other quantities of interest in this work are given by:
\begin{gather*}
Y_k := \frac{\rho_k}{\rho}, \ \X{k} \ := \ \rho_k / M_k, \ c_{v, k} = \bigg( \frac{\partial e_{k}}{\partial T}\bigg), \ c_v := \sum_{k=1}^{N_{s}} Y_k c_{v,k}, \ r \ := \ \sum_{k=1}^{N_{s}} Y_k r_k, \ \gamma := \frac{c_v \ + \ r}{c_v}.
\end{gather*}
$Y_k$ is the species mass fraction, $\X{k}$ is the molar concentration of species $X_k$, $c_v$ is the constant-volume specific heat of the gas mixture, and $\gamma$ is the specific heat ratio. \\
\myindent The PDE system (\ref{eq:PDE_base}) is hyperbolic, i.e for any normal vector $\bf{\hat{n}} = (n_{x}, n_{y})$, the Jacobian of the normal flux $\bf{f}_{n} \ := \ n_{x} \bf{f}_{x} \ + \ n_{y} \bf{f}_{y}$ with respect to $\bf{u}$ is diagonalizable with eigenvalues $(u_n, u_n+a, u_n-a)$ where the normal velocity $u_n$ and speed of sound $a$ are given by:
\begin{equation*}
    u_n \ := \ n_x u \ + \ n_y v, \ a \ := \ \sqrt{\gamma r T}.
\end{equation*}
\subsection{Thermo-Chemistry Model}
\myindent The chemical kinetics temporal source term in (\ref{eq:PDE_base}) has the general form:
\begin{equation*}
      \bf{\Omega} \ := \ \begin{bmatrix}  \big( \Omega_k  \big)_{1 \leq k \leq N_{s}} & 0 & 0 & 0 \end{bmatrix}^T.
\end{equation*}
For standard kinetic models, the source term writes
\begin{equation*}
    \Omega_k \ := \ M_k \sum_{r=1}^{N_r} \ \nu_{r, k} \omega_{r}, \ \omega_{r} \ := \ k_{r}^{f}(T) \prod_{k=1}^{N_s} \X{k}^{\alpha_{r,k}} \ - \ k^{b}_{r}(T) \prod_{k=1}^{N_s} \X{k}^{\beta_{r,k}}.
\end{equation*}
The $(\omega_{r})_{1 \leq r \leq N_{r}}$ terms model $N_{r}$ chemical processes using mass-action kinetics:
\begin{equation}\label{eq:reaction}
    \sum_{k=1}^{N_s} \ \alpha_{r, k} X_k \ \leftrightharpoons \ \sum_{k=1}^{N_s} \ \beta_{r, k} X_k, \ 1 \ \leq \ r \ \leq \ N_{r},
\end{equation}
where $\big( \alpha_{r,k} \big)_{1 \leq k \leq N_{s}}$ and $\big( \beta_{r,k} \big)_{1 \leq k \leq N_{s}}$ denote the sets of stoichiometric coefficients for the reactants and products, respectively, and $\nu_{r,k} \ = \ \beta_{r,k} \ - \ \alpha_{r,k}$. The forward and backward rate coefficients for reaction r are $k_{r}^{f}$ and $k_{r}^{b}$, respectively. They solely depend on temperature and can be related through a chemical equilibrium function $K_{eq,r}(T)$. \\
\myindent For Earth entry \cite{Park_Air}, an important reaction is the dissociation/recombination of diatomic nitrogen:
\begin{equation}\label{eq:N2_diss}
    (DR) \ \ \ N_{2} \ + \ M \ \leftrightharpoons \ 2N \ + \ M, \ M \in \{ N_{2}, N \}. 
\end{equation}
By its endothermic nature, this reaction will consume a portion of the thermal energy generated by the shock. Not accounting for such processes (not just for Earth \cite{Park_Mars}) typically leads to overestimated heating loads, which translates into thermal protection system designs that are not optimal. From a safety perspective however, this prediction needs to be as accurate as possible as it will drive impactful engineering decisions. This context places substantial demands in the fidelity of thermo-chemical models. \\
\myindent The thermo-chemical model we have described so far does not provide reliable results for high-temperature low-density entry configurations, where thermo-chemical non-equilibrium effects at the microscopic level proceed at rates comparable with compressible flow scales \cite{Anderson, Vincenti}. It is known from quantum physics that the internal energy of individual gas particles take discrete values. The distribution of particles among energy levels evolves through collisional and radiative processes and eventually reaches a Boltzmann equilibrium distribution that is solely a function of a few macroscopic quantities. The temperature $T$ is one such parameter, and along the number of particles, it determines the state-resolved population. Representing both the chemical rates $(k_{r}^{f}, k_{r}^{b})$ and the specific internal energies of each species as function of $T$ inherently assume that these processes equilibrium well before macroscopic dissociation/recombination processes such as (DR) take place. \\
\myindent A finer thermo-chemistry model for $N_{2} - N$ dynamics explicitly tracks diatomic nitrogen populations $N_{2}(v)$ among a set number $N_{v}$ of quantum energy levels $\big(e_{v}^{0}\big)_{1 \leq v \leq N_{v}}$, and integrates into the dynamics transition processes between quantum energy states. Such models are referred to as State-to-State (StS) kinetic models  \cite{CapitelliBook} and their integration in CFD simulations has gained some traction over the last decade \cite{Josyula2015, Josyula2017, Bonelli2017, Ninni2022}. We consider the following set of StS processes \cite{Jaffe2010} as a reference system:
\begin{align}
(VTa) \ \ \ \ \ \ \ \ \ N_{2}(v) \ + \ N \ \leftrightharpoons & \ \ N_2(v-\Delta v) \ + \ N, \notag \\
(VTm) \ \ \ \ \ \ \ N_{2}(v) \ + \ N_{2} \ \leftrightharpoons & \ \ N_2(v-1) \ + \ N_{2}, \notag \\
(DRa) \ \ \ \ \ \ \ \ \  N_{2}(v) \ + \ N \ \leftrightharpoons & \ \ 2 N \ + \ N, \label{eq:StS_model} \\
(DRm) \ \ \ \ \ \ \ N_{2}(v) \ + \ N_2 \ \leftrightharpoons & \ \ 2 N \ + \ N_2, \notag \\
(VV)  \ \ \ \ \ N_{2}(v) \ + \ N_2(w) \ \leftrightharpoons & \ \ N_{2}(v-1) \ + \ N_2(v+1). \notag
\end{align}
\myindent At present time, the higher degree of fidelity that StS models can provide CFD efforts is largely offset by its prohibitive costs. The number of species involved is given here by $N_{s} \ = \ N_{v} + 1$ grows with the number of quantum states resolved, which is already large \cite{Jaffe2010}. As will be shown in the next section, the increase in complexity brought about by StS models is grows quadratically with $N_{v}$. \\
\myindent A middle ground between high-fidelity and efficiency that is the current de facto standard in entry simulation consists in using multi-temperature models \cite{Park_book, Candler_review} where different temperatures are introduced to track pseudo equilibrium distributions for distinct internal energy modes. A well-known example is the two-temperature model of Park \cite{Park_2T} where the temperature $T$ is used to represent translational and rotational energy modes, and a \textit{vibrational temperature} $T_v$ is used to represent vibrational and electronic modes. Chemical rates are modeled as functions of both temperatures, and additional energy equations are modeled to evolve the temperatures \cite{Gnoffo1989}. Multi-temperature models can be construed as reduced StS models \cite{Colonna2006, Liu2015, Park_book}. \\
\myindent The quantum chemistry and modeling endeavors involved in putting together StS models such as (\ref{eq:StS_model}) are beyond  the present work. The same goes for lower-level kinetic models accounting for translational non-equilibrium \cite{KustovaBook}.

\section{Implicit Scheme}
\subsection{Solution Representation}
\myindent At this juncture, we introduce the solution vector $\bf{q} \in \mathbb{R}^{m}$ used to discretize the physics. Discrete solutions of the physical model (\ref{eq:PDE_base}) need not be represented only through the temporal flux or conserved variables vector ($\bf{q} := \bf{u}$). Any choice such that the mapping between $\bf{q}$ and $\bf{u}$ is one-to-one is valid and will often be introduced to mitigate stability and algebraic challenges \cite{Gouasmi2022, Turkel, Diosady2015, Chalot1990}. \\
\myindent In this work, we use the primitive variables given by:
\begin{equation*}
    \bf{q} \ := \ \begin{bmatrix} \big( \rho_{k }\big)_{1 \leq k \leq N_{s}} & u & v & T \end{bmatrix}.
\end{equation*}
This representation facilitates Jacobian derivations and solution validity checks (positive densities and temperature). Most notably, it clears out the computational intricacies that stem from not explicitly tracking the temperature \cite{Chalot1990}, which is required to compute pressure and temporal contributions from chemical kinetics. When the internal energy functions are non-trivial functions of temperature, the choice $\bf{q} \ := \bf{u}$ entails cumbersome local nonlinear solves to compute temperature from partial densities and total energy. 
\subsection{Space-Time Discretization}
\myindent The reference discretization is an implicit first-order finite-volume discretization. For each mesh element $\kappa \in \mathcal{K}$, we have:
\begin{equation}\label{eq:fvm_residual}
\bf{u}(\bf{q}_{\kappa}^{n+1}) \ - \ \bf{u}(\bf{q}_{\kappa}^{n}) \ - \ \frac{(\Delta t)_{\kappa}}{V_{\kappa}} \ \int_{\delta \kappa} \bf{\hat{f}}(\bf{q}_{\kappa}^{n+1}, \ \bf{q}_{\kappa'}^{n+1}, \ \bf{\hat{n}}_{\kappa,\kappa'}) \ dS \ = \ \bf{\Omega}(\bf{q}_{\kappa}^{n+1}).
\end{equation}
$\bf{q_{\kappa}^{n}} \in \mathbb{R}^{m}$ denotes the solution vector in element $\kappa$ at discrete time instant $n$. The integral term ($\delta \kappa$ denotes the trace of element $\kappa$) features a numerical flux function $\hat{\bf{f}}$ where $\kappa'$ denotes the neighbor element of $\kappa$ along $\delta \kappa$, and $\bf{\hat{n}}_{\kappa, \kappa'}$ denotes the outward normal vector on $\delta \kappa$. $V_{k}$ denotes the cell size, and $(\Delta t)_k$ denotes a time step applied locally as to accelerate convergence (temporal accuracy is not needed for steady-state solutions). \\
\myindent In the present work, we consider HLL-type numerical fluxes \cite{HLL_flux, HLLE_flux, HLLC_flux} of the form:
\begin{equation}\label{eq:HLL_flux}
    \bf{\hat{f}}(\bf{q}_{L}, \ \bf{q}_{R}, \ \bf{\hat{n}}) \ := \ \frac{s_R \bf{f}_{n, L} \ - \ s_L \bf{f}_{n, R} \ + \ s_L s_R \big( \bf{u}_{R} \ - \ \bf{u}_L\big)}{s_R \ - \ s_L},
\end{equation}
where the subscripts $L$ and $R$ are used to denote generic elements on the left and right sides of an interface, and $\bf{\hat{n}}$ denotes the normal vector oriented from left to right. $s_L$ and $s_R$ are scalar  estimates of the lower and upper bounds of the wave speeds of an equivalent Riemann problem. We are not concerned with the specific way in which $s_L$ and $s_R$ are computed. We simply assume that $s_L$ and $s_R$ are computed from $(u_{n,L}, u_{n,R}, a_{L}, a_{R})$. \\
\myindent Boundary conditions for a typical hypersonic 2D flow past a blunt body at zero angle of attack are:
\begin{itemize}
\item[-] Supersonic inflow/outflow conditions, i.e $ \bf{\hat{f}} \ = \ \bf{f}_{n,L}$, at the beginning (pre-shock) and end (behind the body) of the domain.
\item[-] No-penetration at the wall, i.e $ \bf{\hat{f}} \ = \ \begin{bmatrix} 0 & \dots & 0 & p n_{x} & p n_y & 0 \end{bmatrix}$.
\item[-] Along the symmetry line, $ \bf{\hat{f}} \ = \ \bf{\hat{f}} (\bf{q}_{L}, \ \bf{q}^{mirror}_{L}, \ \bf{\hat{n}})$ where $\bf{q}^{mirror}_{L}$ is a mirror state computed from $\bf{q}_{L}$ and $\bf{\hat{n}}$.
\end{itemize}
\subsection{Nonlinear Solution Method}
\myindent The system (\ref{eq:fvm_residual}) is equivalent to $\bf{R}(\bf{Q}) \ = \ 0$, where $\bf{Q} \ := \ \big( \bf{q}_{\kappa}^{n+1} \big)_{\kappa \in \mathcal{K}}$ the residual vector $\bf{R} \ := \big( \bf{R}_{\kappa}\big)_{k \in \mathcal{K}}$ is given by:
\begin{equation}\label{eq:system_residual}
    \bf{R}_{\kappa}(\bf{Q}) \ := \  \bf{u}(\bf{q}_{\kappa}^{n+1}) \ - \ \bf{u}(\bf{q}_{\kappa}^{n}) \ - \ \frac{(\Delta t)_{\kappa}}{V_{\kappa}} \ \int_{\delta \kappa} \bf{\hat{f}}(\bf{q}_{\kappa}^{n+1}, \ \bf{q}_{\kappa'}^{n+1}, \ \bf{\hat{n}}_{\kappa,\kappa'}) \ dS \ - \ (\Delta t)_{\kappa} \bf{\Omega}(\bf{q}_{\kappa}^{n+1}).
\end{equation}
We use the Newton-Raphson method. An approximate solution is sought through successive iterations $\bf{Q}^{p}$ starting from $\bf{Q}^{0} \ := \ \big( \bf{q}_{\kappa}^{n} \big)_{\kappa \in \mathcal{K}}$ following $\bf{Q}^{p+1} \ := \  \bf{Q}^{p} \ + \ \alpha^{p} \big( \Delta \bf{Q}\big)^{p}$, where $\alpha^p$ is a line search coefficient. The vector $\Delta \bf{Q}^{p+1}$ is the solution of the linear system: 
\begin{equation}\label{eq:LinearSystem}
  \bigg(\frac{\partial \bf{R}}{\partial \bf{Q}}\bigg)_{\bf{Q}^{p}} \ \big( \Delta \bf{Q} \big)^{p+1} \ = \ - \bf{R}(\bf{Q}^{p}).
\end{equation}
The linear system (\ref{eq:LinearSystem}) is typically solved using Krylov methods \cite{Saad_book} such as GMRES \cite{Saad_GMRES}. \\
\myindent Remarkably, solving the linear system does not require the explicit assembly and storage of the discrete Jacobian. An implementation of the matrix-vector product is sufficient. Such Jacobian-free approaches \cite{JFNK} are extremely popular and used in many applications, especially when the size of the system is very large. Jacobian-free approaches have the merit of a lower memory footprint, which is appreciated for full-scale applications. However, the memory savings come at the expense of faster operations. Linear systems are ubiquitous in computational science and there are now several libraries providing high-performance linear algebra capability for various programming languages and various hardware \cite{PETSC, Scipy, CuSPARSE, PyTorch}. Besides enabling optimal compute power, working out the structure of the Jacobian has often brought about significant advances in preconditioning techniques \cite{Pulliam, Turkel, Diosady}.
\subsection{Jacobian Complexity}
\begin{figure}[t!]
    \centering
    \begin{subfigure}[t]{0.32 \textwidth}
      \centering
      \includegraphics[scale=0.55]{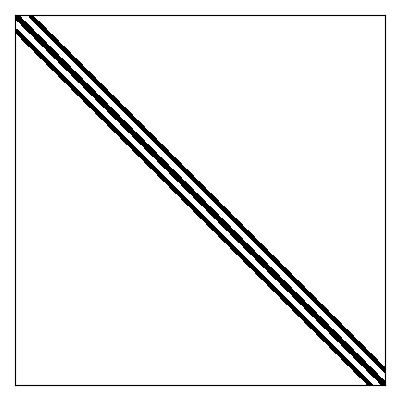}
      \caption{$m=5$}
    \end{subfigure}
    ~
    \begin{subfigure}[t]{0.32 \textwidth}
      \centering
      \includegraphics[scale=0.55]{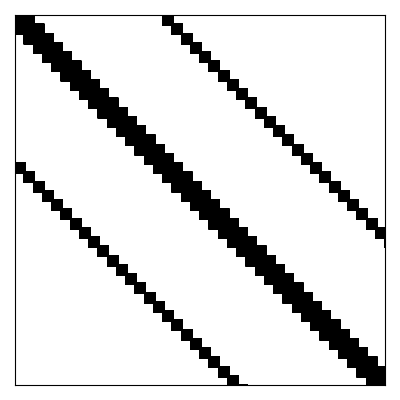}
      \caption{$m=50$}
    \end{subfigure}
    ~
    \begin{subfigure}[t]{0.32 \textwidth}
      \centering
      \includegraphics[scale=0.55]{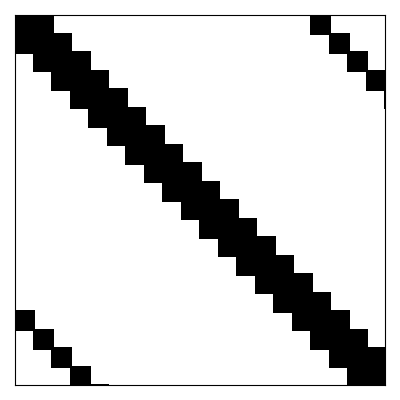}
      \caption{$m=100$}
    \end{subfigure}
    \caption{First $2000 \times 2000$ block of the discrete Jacobian associated with the implicit scheme on a sample $16 \times 32$ structured grid for varying values of $m$ (number of governing equations).}
    \label{fig:bsr_pattern}
\end{figure}

\myindent In the present work, we are interested in the explicit representation and operation of the Jacobian matrix. A standard way of representing $A$ is as a block-sparse matrix. For the first-order discretization we are considering, each Jacobian block is $m \times m$, and the number of non-zero blocks along the row associated with element $\kappa$, is determined by the number of interior edges $N_{\kappa, e}$ of that element. The total number $N_{A}$ of non-zero elements on the standard representation of $A$ can be written as:
\begin{equation*}
    N_{A} \ = \ \sum_{\kappa \in \mathcal{K}} \ \big( m \times m \big) \big( N_{\kappa, e} + 1 \big).
\end{equation*}
\myindent This number, representative of both spatial (storage) and temporal (matrix-vector and matrix-matrix operations) complexity grows quadratically with $m$, which grows with the number of species $N_{s}$. With StS kinetic systems of size ranging anywhere from $46$ quantum levels \cite{Armenise1996}, to just over a hundred \cite{Bonelli2017, Ninni2022}, to over 9000 \cite{Jaffe2010} and more, the complexity increase is considerable. Sample Jacobian patterns are shown in figures \ref{fig:bsr_pattern}(a), \ref{fig:bsr_pattern}(b), and \ref{fig:bsr_pattern}(c). \\
\myindent In the following sections, we describe an alternative representation of the Jacobian whose complexity scales linearly instead of quadratically.

\section{R1-Sparsity}
\myindent In this section, we demonstrate that a more efficient representation of the discrete Jacobian is possible. Let's examine the primary contributions from the inviscid flux and from source term kinetics.
\subsection{Inviscid Flux Jacobian}
\begin{figure}[t!]
    \centering
    \begin{subfigure}[t]{0.49 \textwidth}
      \centering
      \includegraphics[scale=0.65]{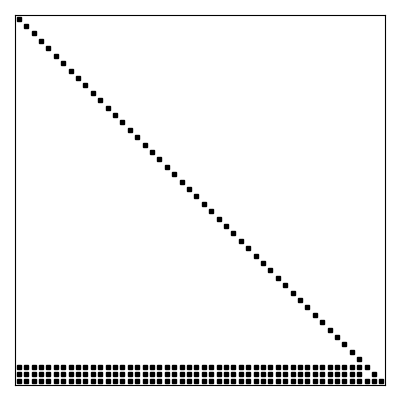}
      \caption{$\bf{H}$}
    \end{subfigure}
    ~
    \begin{subfigure}[t]{0.49 \textwidth}
      \centering
      \includegraphics[scale=0.65]{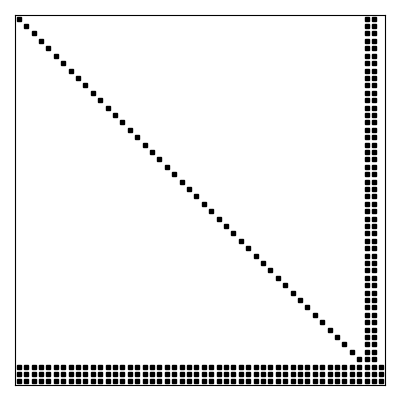}
      \caption{$\bf{A}_{n}$}
    \end{subfigure}
    \caption{Sparsity patterns of the temporal and normal flux Jacobians for $m = 50$.}
    \label{fig:flux_pattern}
\end{figure}

\myindent We are interested in the local Jacobian contributions of $\bf{\hat{f}} \ = \ \bf{\hat{f}}(\bf{q}_{L}, \ \bf{q}_{R}, \  \bf{\hat{n}})$ with respect to the left and right states. Let's define:
\begin{equation*}
    \bf{\hat{A}}_{ L} \ := \ \bigg(\frac{\partial \bf{\hat{f}}}{\partial \bf{q}_L}\bigg) \ \mbox{and} \  \bf{\hat{A}}_{R} \ := \ \bigg(\frac{\partial \bf{\hat{f}}}{\partial \bf{q}_R}\bigg).
\end{equation*}
We can rewrite equation (\ref{eq:HLL_flux}) as:
\begin{equation}\label{eq:HLL_flux_compact}
    \bf{\hat{f}}(\bf{q}_{L}, \ \bf{q}_{R}, \ \bf{\hat{n}}) \ := \ \frac{\bf{f}_{n,L} \ + \ \bf{f}_{n, R}}{2} \ - \ a_{f} \big( \bf{f}_{n, R} \ - \ \bf{f}_{n, L} \big) \ + \ c_{f} \big( \bf{u}_{R} \ - \ \bf{u}_{L} \big).
\end{equation}
where the scalar coefficients $a_{f}$ and $c_{f}$ are given by:
\begin{equation*}
    a_{f} \ := \ \frac{1}{2} \frac{s_L \ + \ s_R}{s_R \ - \ s_L}, \ c_{f} \ := \ \frac{s_L s_R}{s_R \ - \ s_L}.
\end{equation*}
It can be easily shown that the local Jacobians: 
\begin{equation*}
    \bf{A}_{n} \ := \ \bigg( \frac{\partial \bf{f}(\bf{q}, \bf{\hat{n}})}{\partial \bf{q}}\bigg) \ \mbox{and} \ \bf{H} \ := \ \bigg( \frac{\partial \bf{u}}{\partial \bf{q}} \bigg),
\end{equation*}
are sparse matrices. This is depicted in figures \ref{fig:flux_pattern}-(a) and \ref{fig:flux_pattern}-(b). \\
\myindent Upon closer examination of equation (\ref{eq:HLL_flux_compact}), it follows that the density of the matrices $\bf{\hat{A}}_{L/R}$ is due to the scalar coefficients $a_{f}$ and $c_{f}$, which depend on all components of $\bf{q}$ since the wave speeds are determined by $(u, v)$ and the speed of sound, which itself depends on the species mass fractions and the temperature. In other words, the dense part of $\bf{\hat{A}}_{L/R}$ can be compactly represented as the sum of two rank-one matrices. We have:
\begin{align}
     \bf{\hat{A}}_{ L}  \ =& \ \ \bf{\hat{A}}_{L}^{0} \ - \ \big( \bf{f}_{n, R} \ - \ \bf{f}_{n, L} \big) \bigg( \frac{\partial a_{f}}{\partial \bf{q}_L}\bigg)  \ + \ \big( \bf{u}_{R} \ - \ \bf{u}_{L} \big) \bigg( \frac{\partial c_{f}}{\partial \bf{q}_L}\bigg), \label{eq:r1_flux_left} \\
     \bf{\hat{A}}_{ R}  \ =& \ \  \bf{\hat{A}}_{R}^{0} \ - \ \big( \bf{f}_{n, R} \ - \ \bf{f}_{n, L} \big) \bigg( \frac{\partial a_{f}}{\partial \bf{q}_R}\bigg)  \ + \ \big( \bf{u}_{R} \ - \ \bf{u}_{L} \big) \bigg( \frac{\partial c_{f}}{\partial \bf{q}_R}\bigg),  \label{eq:r1_flux_right}
\end{align}
where $\bf{\hat{A}}_{L/R}^{0}$ are sparse matrices given by:
\begin{align*}
    \bf{\hat{A}}_{L}^{0} \ := & \ \ \bigg( \frac{1}{2} \ + \ a_{f} \bigg) \bf{A}_{n, L} \ - \ c_f \bf{H}_{L}, \\ \bf{\hat{A}}_{R}^{0} \ :=& \ \ \bigg( \frac{1}{2} \ - \ a_{f} \bigg) \bf{A}_{n, R}\ + \ c_f \bf{H}_{R}.
\end{align*}
Equations (\ref{eq:r1_flux_left}) and (\ref{eq:r1_flux_right}) express the main idea of this work, namely that \textit{some dense matrices can be represented as the sum of a sparse matrix and rank-one updates.}. We refer to such matrices as \textit{r1-sparse} when the number of rank-one updates is negligible compared to base matrix dimension $m$. Under this condition, we can expect the desired linear spatio-temporal complexity with respect to $m$.
\begin{figure}[t!]
    \centering
    \includegraphics[scale=0.5]{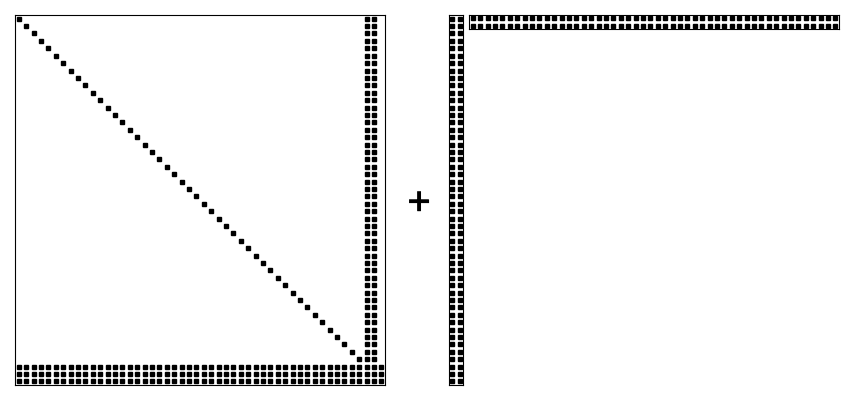}
    \caption{r1-sparse representation of the local trace flux Jacobian $\bf{\hat{A}}$ for $m = 50$.}
    \label{fig:r1_sparse_trace_flux}
\end{figure}
\subsection{Temporal Source Jacobian}
\begin{figure}[t!]
    \centering
    \begin{subfigure}[t]{0.49 \textwidth}
      \centering
      \includegraphics[scale=0.7]{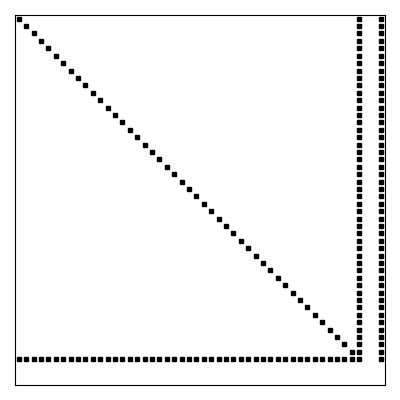}
      \caption{$(DRa)$ process.}
    \end{subfigure}
    ~
    \centering
    \begin{subfigure}[t]{0.49 \textwidth}
      \centering
      \includegraphics[scale=0.7]{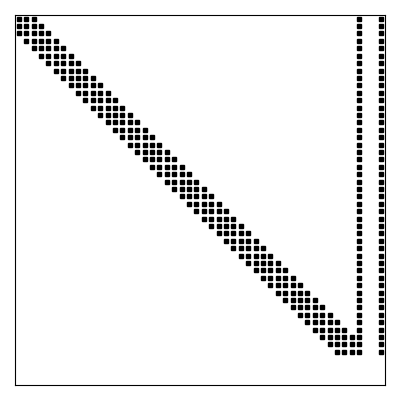}
      \caption{$(VTa)$ process.}
    \end{subfigure}
    ~
    \begin{subfigure}[t]{0.5 \textwidth}
      \centering
      \includegraphics[scale=0.7]{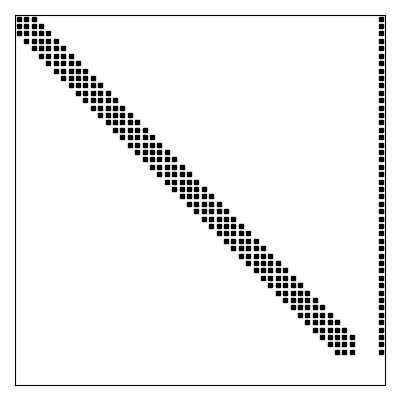}
      \caption{$(VV)$ process.}
    \end{subfigure}
    \caption{Sparsity patterns of temporal source contributions for $m=50$ ($46$ energy levels of $N_{2}$). The bandwidth depends on the threshold energy jump $\Delta v$ of the model. }
    \label{fig:temporal_source_pattern}
\end{figure}

\myindent Let's define the local Jacobian:
\begin{equation*}
    \bf{G} \ := \ \bigg( \frac{\partial \bf{\Omega}}{\partial \bf{q}} \bigg).
\end{equation*}
For the model $N_{2}-N$ StS kinetic system (\ref{eq:StS_model}), the source term $\bf{\Omega}$ writes as a sum of contributions from the various state-resolved processes outlined in section 2. We can write:
\begin{align*}
    \bf{\Omega} (\bf{q}) \ :=& \ \bf{\Omega}^{(VTa)}(\bf{q}) \ + \ \bf{\Omega}^{(VTm)}(\bf{q}) \ + \  \bf{\Omega}^{(DRa)}(\bf{q}) \ + \ \bf{\Omega}^{(DRm)}(\bf{q}) \ + \ \bf{\Omega}^{(VV)}(\bf{q}), \\
    \bf{G}(\bf{q})  \ :=& \ \bf{G}^{(VTa)}(\bf{q}) \ + \ \bf{G}^{(VTm)}(\bf{q}) \ + \  \bf{G}^{(DRa)}(\bf{q}) \ + \ \bf{G}^{(DRm)}(\bf{q}) \ + \ \bf{G}^{(VV)}(\bf{q}).
\end{align*}
In the same spirit as in the previous subsection, we can examine each contribution individually. \\
\myindent The contribution from $DRa$ processes $N_{2}(v) \ + \ N \ \leftrightharpoons \ 2 N \ + \ N$ is sparse. For a given quantum level $v$, there are three non-zero entries in the corresponding Jacobian row (sensitivity with respect to the partial densities of $N_2(v)$ and $N$, and with respect to temperature $T$). A sample sparsity pattern is shown in figure (\ref{fig:temporal_source_pattern})-(a). \\
\myindent The contribution from $VTa$ processes $N_{2}(v) \ + \ N \ \leftrightharpoons \ N_2(v - \Delta v) \ + \ N$ is sparse to the extent that processes with $|\Delta v|$ larger than some threshold value can be neglected depending on the considered temperature range \cite{Armenise1996}. A similar argument holds for $VV$ processes $N_{2}(v) \ + \ N_{2}(w) \ \leftrightharpoons \ N_2(v - 1) \ + \ N_{2}(w+1)$ and $|(v-w)|$. Sample sparsity patterns are shown in figures (\ref{fig:temporal_source_pattern})-(b) and (\ref{fig:temporal_source_pattern})-(c).\\
\myindent The density of the Jacobian of $\bf{\Omega}$ comes from $DRm$ and $VTm$ contributions. These processes are sensitive to all species since $N_{2}$ can be any of $\big(N_{2}(v) \big)_{1 \leq v \leq N_{v}}$. Fortunately, these contributions can be represented as single rank-one updates as well. We can write:
\begin{align*}
   \bf{G}^{(DRm)} \ =& \ \bigg( \frac{\partial \bf{\bar{\Omega}}^{(DRm)}}{\partial \bf{q}} \bigg) \rho_{N_{2}} \ + \ \bf{\bar{\Omega}}^{(DRm)} \bigg( \frac{\partial \rho_{N_{2}}}{\partial \bf{q}}\bigg), \\
    \bf{G}^{(VTm)}  \ =& \ \bigg( \frac{\partial \bf{\bar{\Omega}}^{(VTm)}}{\partial \bf{q}} \bigg) \rho_{N_{2}} \ + \ \bf{\bar{\Omega}}^{(VTm)} \bigg( \frac{\partial \rho_{N_{2}}}{\partial \bf{q}}\bigg).
\end{align*}
where $\bf{\Omega}^{(DRm)} \ := \bf{\bar{\Omega}}^{(DRm)} \rho_{N_{2}}$ and $\bf{\Omega}^{(VTm)} \ := \bf{\bar{\Omega}}^{(VTm)} \rho_{N_{2}}$. Therefore, the r1-sparse representation of $\bf{G}$ is given by:
\begin{equation}\label{eq:r1_chem}
    \bf{G} \ = \ \bf{G}^{0} \ + \ \bigg( \bf{\bar{\Omega}}^{(DRm)}(\bf{q}) \ + \ \bf{\bar{\Omega}}^{(VTm)}(\bf{q})\bigg) \bigg( \frac{\partial \rho_{N_{2}}}{\partial \bf{q}}\bigg),
\end{equation}
that is one rank-one update and a sparse contribution $\bf{G}^{0}$ given by:
\begin{equation*}
    \bf{G}^{0} \ = \  \bf{G}^{(VTa)} \ + \  \bf{G}^{(DRa)} \ + \ \bf{G}^{(VV)} \ + \ \bigg[  \bigg( \frac{\partial \bf{\bar{\Omega}}^{(DRm)}}{\partial \bf{q}} \bigg) \ + \ \bigg( \frac{\partial \bf{\bar{\Omega}}^{(VTm)}}{\partial \bf{q}} \bigg) \bigg] \rho_{N_{2}}.
\end{equation*}
\begin{figure}[htbp!]
    \centering
    \includegraphics[scale=0.5]{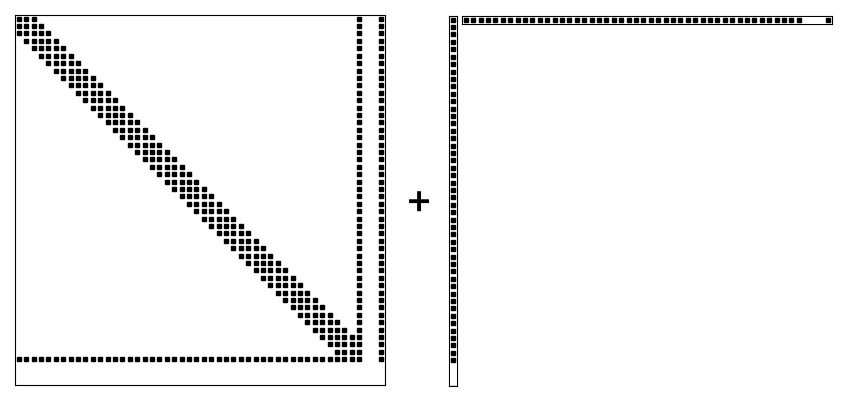}
    \caption{r1-sparse representation of the local temporal source Jacobian $\bf{G}$ for $m = 50$.}
    \label{fig:r1_sparse_source}
\end{figure}
\subsection{Overall Structure}
\begin{figure}[htbp!]
    \centering
    \includegraphics[scale=0.5]{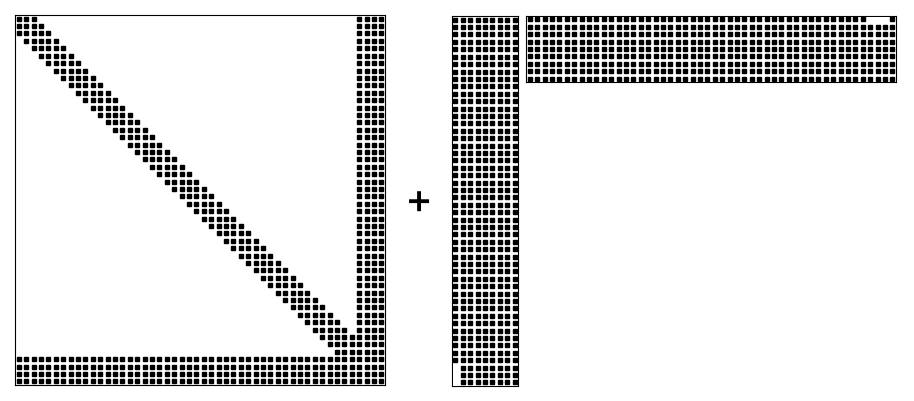}
    \caption{r1-sparse representation of a Jacobian's diagonal block for $m=50$. The sparsity pattern blends in those shown in figures \ref{fig:flux_pattern}(a)-(b), \ref{fig:temporal_source_pattern}(a)-(c), and \ref{fig:r1_sparse_trace_flux}. }
    \label{fig:r1_sparse_core}
\end{figure}

\begin{figure}[t!]
    \centering
    \begin{subfigure}[t]{0.49 \textwidth}
      \centering
      \includegraphics[scale=0.6]{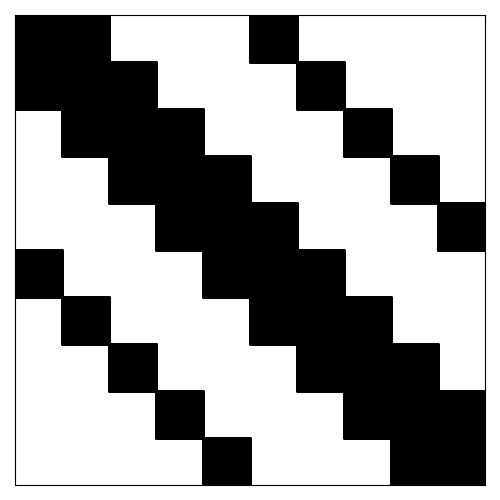}
      \caption{Standard representation}
    \end{subfigure}
    ~
    \centering
    \begin{subfigure}[t]{0.49 \textwidth}
      \centering
      \includegraphics[scale=0.6]{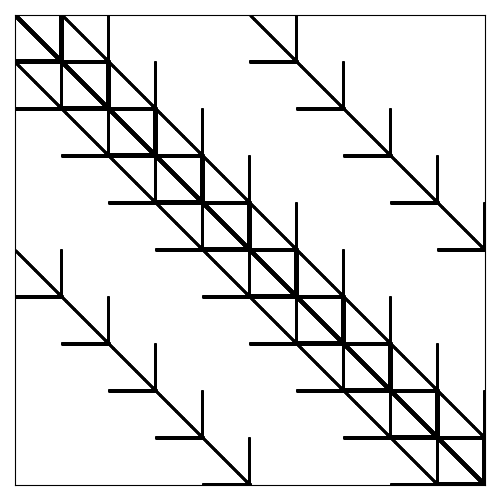}
      \caption{Fully sparse contribution.}
    \end{subfigure}
    ~
    \begin{subfigure}[t]{0.49 \textwidth}
      \centering
      \includegraphics[scale=0.6]{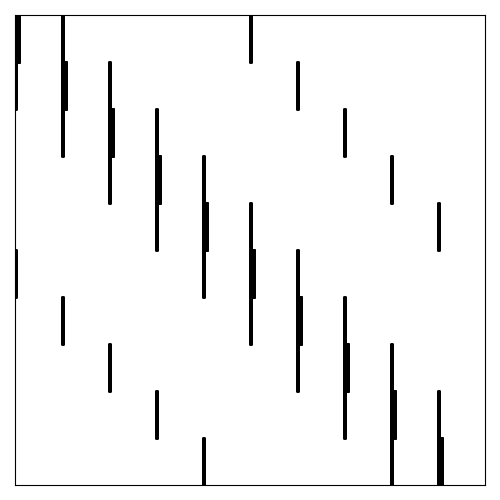}
      \caption{Block-wise column matrix.}
    \end{subfigure}
    \begin{subfigure}[t]{0.49 \textwidth}
      \centering
      \includegraphics[scale=0.6]{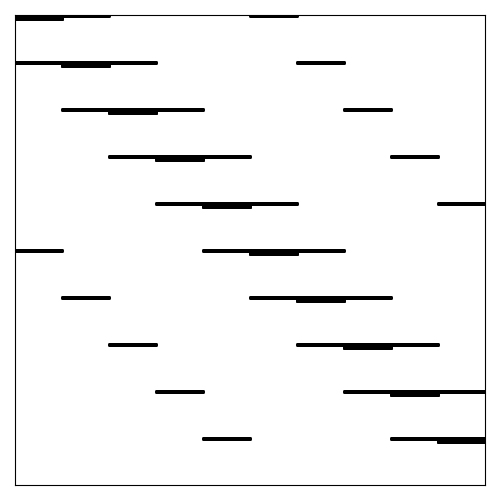}
      \caption{Block-wise row matrix.}
    \end{subfigure}
    \caption{Two representations of the discrete Jacobian for $m=100$ on a $5 \times 2$ structured grid. The standard representation is shown in (a). Our alternative combines a fully-sparse matrix (b) with local rank-one updates where all local column vectors, shown in (c), and all local row vectors, shown in (d), are stored in two distinct data structures. Note the differences in local sparsity and number of local row/columns between diagonal blocks and off-diagonal blocks.}
    \label{fig:final_r1_sparse}
\end{figure}

\myindent Figure \ref{fig:final_r1_sparse} shows the representation that comes forth when dense Jacobian blocks are represented using r1-sparse matrices. Off-diagonal Jacobian blocks are from interior trace flux contributions, hence their r1-sparsity pattern follows that of figure \ref{fig:r1_sparse_trace_flux}. The r1-sparsity pattern of Jacobian blocks on the diagonal, shown in figure \ref{fig:r1_sparse_core}, is obtained by merging the respective patterns of the temporal flux, the temporal source, and all local trace contributions. \\
\myindent Note that the local r1-sparse representation along the diagonal contains $9$ local rank-one updates. One local rank update comes from the temporal source term (StS kinetics). The remaining eight are the product of up to four trace flux contributions bringing each 2 local rank-one updates. The number four is the maximum number of neighboring elements in the 2D grid.

\section{Implementation}
\myindent The goal of this section is to introduce working storage formats and implementations of block r1-sparse matrices that achieve linear complexity in $N_{s}$. The code listings are written in the python programming language for simplicity, and can be adapted to more efficient languages such as C++. The optimization of these routines on various hardware \cite{Wellein_book} is the topic of future work. 
\subsection{Sparse Storage Formats}
\myindent One of the most popular storage formats for sparse matrices is the Compressed Sparse Row (CSR) format \cite{Saad_book, Golubook}. Along with basic information (shape and data type), it consists of three one-dimensional arrays:
\begin{enumerate}
    \item A \texttt{data} array containing all the non-zero matrix entries.
    \item An \texttt{indices} array containing the column indices of all the non-zero matrix entries.
    \item An \texttt{indptr} array of index pointers to \texttt{indices} that help locate the non-zero entries of a set matrix row. This array has as many entries as the number of rows plus one.
\end{enumerate}
For example, the matrix:
\begin{equation*}
    \begin{bmatrix}
        11 & 0 & 0 & -1 \\
        2 & 3 & 0 & 0 \\
        0 & 0 & 23 & 19 \\
        -1 & -2 & -3 & - 4 
    \end{bmatrix}
\end{equation*}
has the following CSR representation.
\begin{align*}    
    \texttt{data} \ =& \ \begin{bmatrix} 11 & -1 & 2 & 3 & 23 & 19 & -1 & -2 & -3 & -4 \end{bmatrix}, \\
    \texttt{indices} \ =& \ \begin{bmatrix} 0 & 3 & 0 & 1 & 2 & 3 & 0 & 1 & 2 & 3 \end{bmatrix}, \\
    \texttt{indptr} \ =& \ \begin{bmatrix} 0 & 2 & 4 & 6  & 10 \end{bmatrix}.
\end{align*}
An equivalent representation is the Compressed Sparse Column (CSC) format for matrices stored in column-major order. Different storage formats serve different purposes (facilitating sparse array creation and optimizing specific linear algebra operations). Standard libraries such as Scipy \cite{Scipy} implement most of them, and can easily convert between formats as seen fit. \\
\myindent An r1-sparse matrix can be represented with CSR/CSC data combined with \texttt{r1\_columns} and \texttt{r1\_rows} two-dimensional arrays (first dimension being the number of rank-one updates, second dimension being the length of each column/row). For example, the above matrix can be combined with two rank-one updates:
\begin{equation*}
    \begin{bmatrix}
        12 & 1 \\
        -1 & 5 \\
        0 & -2 \\
        1 & 1
    \end{bmatrix} \begin{bmatrix}
        15 & 0 & 30 & 2.5 \\
        -3 & -2.5 & 60 & 0.5
    \end{bmatrix}
\end{equation*}
represented as:
\begin{equation*}
    \texttt{r1\_columns} \ = \ \begin{bmatrix}
        12 & -1 & 0 & 1 \\
        1 & 5 & -2 & 1
    \end{bmatrix}, \
    \texttt{r1\_rows} \ = \ \begin{bmatrix}
        15 & 0 & 30 & 2.5 \\
        -3 & -2.5 & 60 & 0.5
    \end{bmatrix}.
\end{equation*}
\myindent For block-sparse matrices such as the one of interest in this work, the Block Sparse Row (BSR) format is convenient for assembly. It mimics the CSR format except that the \texttt{data} array is now three dimensional and that the \texttt{indices} and \texttt{indptr} contain to block indices (we rename them \texttt{block\_indices} and \texttt{block\_indptr} for clarity). The first dimension of \texttt{data} is the number of non-zero matrix blocks, and the remaining two dimensions correspond to those of each block. For example, the matrix:
\begin{equation*}
    \begin{bmatrix}
        11 & 12  & 0  & 0   & 0 & 0 & 15 & 0 \\
        0 & 14  & 0  & 0   & 0 & 0 & 17 & 0 \\
        0 & 0 & 0 & 31 & 0 & 0 & 0   & 0 \\
        23 & 24 & 32 & 0 & 0 & 0 & 0   & 0 \\
        0   & 0   &   0 & 0 & -7 & -8 & -13 & 0 \\
        0   & 0   &   0 & 0 & -9 & -11 & 0 & -16\\
        0 & 0 & 1 & 0  &  0 & 0 & 7 & 0 \\
        0 & 0 & 9  & 10 & 0 & 0 & 0 & 16 \\
    \end{bmatrix}
\end{equation*}
has the following representation:
\begin{align*}
    \texttt{block\_indices} \ =& \ \begin{bmatrix} 0 & 3 & 0 & 1 & 2 & 3 & 1 & 3 \end{bmatrix}, \\
    \texttt{block\_indptr} \ =& \ \begin{bmatrix} 0 & 2 & 4 & 6 & 8\end{bmatrix}, \\
    \texttt{data} \ =& \ \begin{bmatrix}
    \begin{bmatrix} 11 & 12 \\ 0 & 14 \end{bmatrix} & \! \! \! \!
    \begin{bmatrix} 15 & 0 \\ 17 & 0 \end{bmatrix} & \! \! \! \!
    \begin{bmatrix} 0 & 0 \\ 23 & 24 \end{bmatrix} & \! \! \! \!
    \begin{bmatrix} 0 & 31 \\ 32 & 0 \end{bmatrix} & \! \! \! \!
    \begin{bmatrix} -7 & -8 \\ -9 & -11 \end{bmatrix} & \! \! \! \!
    \begin{bmatrix} -13 & 0 \\ 0 & -16 \end{bmatrix} & \! \! \! \!
    \begin{bmatrix} 1 & 0 \\ 9 & 10 \end{bmatrix} & \! \! \! \!
    \begin{bmatrix} 7 & 0 \\ 0 & 16 \end{bmatrix}
    \end{bmatrix}.
\end{align*}
In our context, the \texttt{indices} and \texttt{indptr} arrays are entirely determined by the adjacency matrix of the mesh. The dimensions of each local block is determined by the number of governing equations $m$. \\
\myindent For a block r1-sparse matrix, the representation of block-wise rank-one contributions is similar to the BSR representation, except that it must account for the fact that the number of rank-one updates is not uniform among non-zero blocks. Consider the following contribution to the BSR matrix above:
\begin{gather*}
    \begin{bmatrix}
        \bf{u}_{0,0} \bf{v}_{0,0}^{T} & \bf{0}_{2 \times 2} & \bf{0}_{2 \times 2} &  \bf{u}_{0,3} \bf{v}_{0,3}^{T} \vspace{0.1 cm} \\
        \bf{u}_{1,0} \bf{v}_{1,0}^{T} &  \bf{u}_{1,1} \bf{v}_{1,1}^{T}  & \bf{0}_{2 \times 2} & \bf{0}_{2 \times 2} \vspace{0.1 cm} \\
        \bf{0}_{2 \times 2} & \bf{0}_{2 \times 2} & \bf{u}_{2,2} \bf{v}_{2,2}^{T} &  \bf{u}_{2,3} \bf{v}_{2,3}^{T} \vspace{0.1 cm} \\
        \bf{0}_{2 \times 2} & \bf{u}_{3,1} \bf{v}_{3,1}^{T} & \bf{0}_{2 \times 2} & \bf{u}_{3,3} \bf{v}_{3,3}^{T}
    \end{bmatrix}, \ \bf{0}_{2 \times 2} \ := \ \begin{bmatrix} 0 & 0 \\ 0 & 0 \end{bmatrix}, \\ \\
     (\bf{u}_{k,k}, \bf{v}_{k, k}) \in \big( \mathbb{R}^{2 \times 3} \times  \mathbb{R}^{2 \times 3}  \big), \  (\bf{u}_{k,j}, \bf{v}_{k, j}) \in \big( \mathbb{R}^{2 \times 1} \times  \mathbb{R}^{2 \times 1}  \big), \ k \neq j.
\end{gather*}
In this example, the number of local rank-one pairs is 3 for diagonal blocks and 1 for off-diagonal blocks. We refer to such matrices as BR1 matrices. A complete working representation is given by:
\begin{align*}
    \texttt{block\_indices} \ =& \ \begin{bmatrix} 0 & 3 & 0 & 1 & 2 & 3 & 1 & 3 \end{bmatrix}, \\
    \texttt{block\_indptr} \ =& \ \begin{bmatrix} 0 & 2 & 4 & 6 & 8 \end{bmatrix}, \\
    \texttt{column\_data} \ =& \ \begin{bmatrix}
    \bf{u}_{0,0} & \bf{u}_{0,3} & \bf{u}_{1,0} & \bf{u}_{1,1} & \bf{u}_{2,2} & \bf{u}_{2,3} & \bf{u}_{3,1} & \bf{u}_{3,3} \end{bmatrix}^{T}, \\
    \texttt{row\_data} \ =& \ \begin{bmatrix}
    \bf{v}_{0,0} & \bf{v}_{0,3} & \bf{v}_{1,0} & \bf{v}_{1,1} & \bf{v}_{2,2} & \bf{v}_{2,3} & \bf{v}_{3,1} & \bf{v}_{3,3} \end{bmatrix}^{T}, \\
    \texttt{pair\_indptr} \ =& \ \begin{bmatrix} 0 & 3 & 4 & 5 & 8 & 11 & 12 & 13 & 16 \end{bmatrix},
\end{align*}
where the \texttt{pair\_indptr} array fulfills a similar role to \texttt{block\_indptr} by providing the starting and end indices in the \texttt{column\_data} / \texttt{row\_data} to access all rank-one pairs located within a certain non-zero block. We call this representation the BR1-BSR format. Additional index arrays can be introduced to optimize linear algebra operations. For example, the BR1 matrix-vector product implementation presented in the next subsection uses an auxiliary pair index pointer array
\begin{equation*}
    \texttt{column\_pair\_indptr} \ = \ \begin{bmatrix} 0 & 4 & 8 & 12 & 16 \end{bmatrix}
\end{equation*} 
that helps access all \texttt{column\_data} entries located on a given row of blocks.
\subsection{Jacobian-Vector Product}
\begin{figure}[h!]
    \centering
    \includegraphics[scale=0.75]{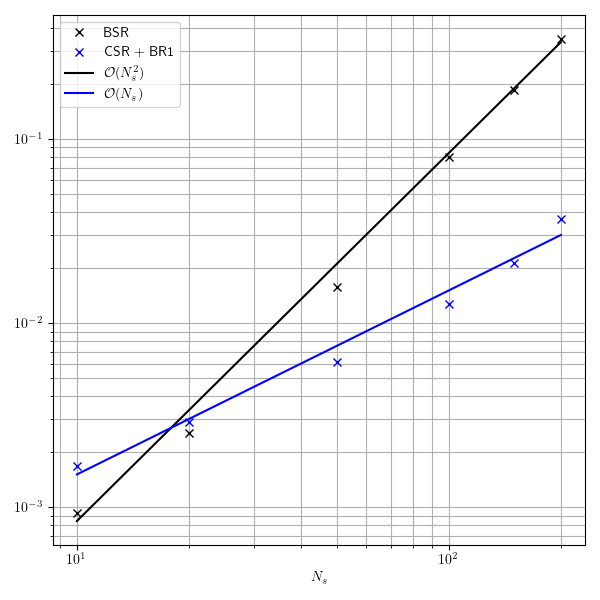}
    \caption{CPU timings for Jacobian-vector product implementations for a structured $40 \times 40$ grid for various values of $N_{s}$.}
    \label{fig:mat_vec_timings}
\end{figure}

\myindent A working python implementation of the matrix-vector product for a CSR matrix is provided in Listing \ref{lst:csr_mat_vec} for reference. The implementation for BSR matrices uses the same routine after conversion to CSR format.
\begin{lstlisting}[
    language=Python,
    frame=tb,
    caption=Sample CSR matrix-vector product implementation,
    captionpos=t,
    label={lst:csr_mat_vec},
]
import numpy as np
from numba import njit

@njit
def csr_matrix_vector_product(
    data: np.ndarray,
    indices: np.ndarray,
    indptr: np.ndarray,
    input_vector: np.ndarray,
    output_vector: np.ndarray,
) -> None:
    nb_rows = indptr.shape[0] - 1
    for row_id in range(nb_rows):
    
        dot_product = 0.
        
        row_start, row_end = indptr[row_id:row_id + 2]
        for data_index in range(row_start, row_end):
            column_index = indices[data_index]
            dot_product += (
               data[data_index] * input_vector[column_index]
            )
        
        output_vector[row_id] = dot_product
\end{lstlisting}
\myindent A working python implementation of the matrix-vector product involving BR1 matrices discussed in section 5.1 is provided in Listing \ref{lst:br1_mat_vec}. For the example BR1 matrix introduced in the section 5.1, we have:
\begin{equation*}
    \begin{bmatrix}
        \bf{u}_{0,0} \bf{v}_{0,0}^{T} & \bf{0}_{2 \times 2} & \bf{0}_{2 \times 2} &  \bf{u}_{0,3} \bf{v}_{0,3}^{T} \vspace{0.1 cm} \\
        \bf{u}_{1,0} \bf{v}_{1,0}^{T} &  \bf{u}_{1,1} \bf{v}_{1,1}^{T}  & \bf{0}_{2 \times 2} & \bf{0}_{2 \times 2} \vspace{0.1 cm} \\
        \bf{0}_{2 \times 2} & \bf{0}_{2 \times 2} & \bf{u}_{2,2} \bf{v}_{2,2}^{T} &  \bf{u}_{2,3} \bf{v}_{2,3}^{T} \vspace{0.1 cm} \\
        \bf{0}_{2 \times 2} & \bf{u}_{3,1} \bf{v}_{3,1}^{T} & \bf{0}_{2 \times 2} & \bf{u}_{3,3} \bf{v}_{3,3}^{T}
    \end{bmatrix} \begin{bmatrix}
       \bf{x}_{0} \vspace{0.1 cm} \\
       \bf{x}_{1} \vspace{0.1 cm} \\
       \bf{x}_{2} \vspace{0.1 cm} \\
       \bf{x}_{3} \end{bmatrix} \ = \ \begin{bmatrix}
        \bf{u}_{0,0} \big( \bf{v}_{0,0}^{T}  \bf{x}_{0} \big) + \bf{u}_{0,3} \big( \bf{v}_{0,3}^{T}  \bf{x}_{3} \big) \vspace{0.1 cm} \\
       \bf{u}_{1,0} \big( \bf{v}_{1,0}^{T}  \bf{x}_{0} \big) + \bf{u}_{1,1} \big( \bf{v}_{1,1}^{T}  \bf{x}_{1} \big)  \vspace{0.1 cm} \\
       \bf{u}_{2, 2}  \big( \bf{v}_{2,2}^{T}  \bf{x}_{2} \big) + \bf{u}_{2, 3}  \big( \bf{v}_{2,3}^{T}  \bf{x}_{3} \big)  \vspace{0.1 cm} \\
        \bf{u}_{3, 1}  \big( \bf{v}_{3,1}^{T}  \bf{x}_{1} \big) + \bf{u}_{3, 3}  \big( \bf{v}_{3,3}^{T}  \bf{x}_{3} \big)  \end{bmatrix}.
\end{equation*}
This implementation involves a buffer array \texttt{column\_coeffs} which stores the intermediate result:
\begin{equation*}
    \texttt{column\_coeffs} \ = \begin{bmatrix} \bf{v}_{0,0}^{T}  \bf{x}_{0} & \bf{v}_{0,3}^{T}  \bf{x}_{3} &  \bf{v}_{1,0}^{T}  \bf{x}_{0} &  \bf{v}_{1,1}^{T}  \bf{x}_{1} & \bf{v}_{2,2}^{T}  \bf{x}_{2} & \bf{v}_{2,3}^{T}  \bf{x}_{3} & \bf{v}_{3,1}^{T}  \bf{x}_{1} & \bf{v}_{3,3}^{T}  \bf{x}_{3} \end{bmatrix}.
\end{equation*}
\begin{lstlisting}[
    language=Python,
    frame=tb,
    caption=BR1 sparse matrix-vector product implementation,
    captionpos=t,
    label={lst:br1_mat_vec},
]
@njit
def br1_mat_vec(
    row_data: np.ndarray,
    block_indices: np.ndarray,
    block_indptr: np.ndarray,
    pair_indptr: np.ndarray,
    column_coeffs: np.ndarray,
    column_data: np.ndarray,
    column_pair_indptr: np.ndarray,
    input_vector: np.ndarray,
    output_vector: np.ndarray,
    column_size: int,
    row_size: int,
) -> None:
    nb_blocks = len(block_indptr) - 1

    for row_block_id in range(nb_blocks):

        # Compute column coeffs for the entire row
        block_start, block_end = block_indptr[row_block_id:row_block_id+2]

        for block_index in range(block_start, block_end):

            # Fetch nonzero chunk
            column_block_id = block_indices[block_index]

            # Fetch input sub-vector
            input_start = column_block_id*row_size

            # Fetch number of pairs
            pair_start, pair_end = pair_indptr[block_index:block_index+2] 

            for pair_index in range(pair_start, pair_end):
                acc = 0.
                for k in range(row_size):
                    acc += (
                        row_data[pair_index, k] * input_vector[input_start + k]
                    )
                column_coeffs[pair_index] = acc
        
        # Complete
        row_start = row_block_id * column_size
        row_end = row_start + column_size

        output_column = output_vector[row_start:row_end]

        pair_start, pair_end = column_pair_indptr[row_block_id:row_block_id+2]
        for pair_index in range(pair_start, pair_end):
            col_coeff = column_coeffs[pair_index]
            for k in range(column_size):
                output_column[k] += col_coeff * column_data[pair_index, k]
\end{lstlisting}
\myindent CPU timings are shown in figure \ref{fig:mat_vec_timings} where we compare two equivalent Jacobian representations for a fixed grid and increasing the number of species $N_{s}$ (equivalently $m$):
\begin{enumerate}
    \item The standard one using BSR format. We used SciPy's implementation of the matrix-vector product (\texttt{scipy.sparse.bsr\_array.dot}).
    \item The new one we developed in this work. It consists of a fully-sparse part which we represented as a CSR matrix and a BR1 matrix. The matrix-vector product is computed by adding up the CSR product (\texttt{scipy.sparse.csr\_array.dot}) and the BR1 product (Listing \ref{lst:br1_mat_vec}).
\end{enumerate}
As expected, the cost of the standard representation grows quadratically whereas ours grows linearly.

\section{A Variation of the Block-Jacobi Preconditioner}
\myindent Solutions of linear systems of the form:
\begin{equation}\label{eq:linear_system}
\bf{Ax \ = \ b}, \ \bf{A} \in \mathbb{R}^{N \times N}, \ (\bf{x}, \bf{b}) \in \mathbb{R}^{N} \times \mathbb{R}^{N},
\end{equation}
using iterative methods such as GMRES \cite{Saad_GMRES} can be accelerated using preconditioning techniques \cite{Saad_book}. This amounts to introducing an invertible operator $\bf{M}$ to modify the original system (\ref{eq:linear_system}). Right-preconditioning techniques compute a solution $\bf{x}$ in two steps:
\begin{equation} \label{eq:right_precond_system}
\bf{x \ = \ M y, \ (AM) y \ = b.}
\end{equation}
Left-preconditioning techniques compute $\bf{x}$ in as the solution of:
\begin{equation} \label{eq:left_precond_system}
\bf{(M A) x \ = \ M b.}
\end{equation}
For efficiency, the operator $\bf{M}$ should be such that the modified operator $\bf{AM}$ (or $\bf{MA}$) has a better condition number and auxiliary operations such as $\bf{My}$ are fast enough. Most preconditioners can be construed as some approximation of the inverse of $\bf{A}$. \\
\myindent When the matrix $\bf{A}$ possesses a square block-structure,  i.e. $N = m \times n_{K}$ where $n_{K}$ is the number of blocks (mesh elements) and $m$ is number of local rows/columns within each block (number of governing equations), a popular choice for $\bf{M}$ is given by the inverse of the block-diagonal component of $\bf{A}$. This preconditioner is commonly referred to as block-Jacobi. Due to its block-diagonal structure, the action of $\bf{M}$ on a vector can be done in parallel, block-by-block. The inverse of $\bf{M}$ is typically represented using local LU factorizations for each diagonal block \cite{Saad_book, Golubook} of $\bf{A}$. \\
\myindent In the context of the present work, we can write each diagonal block $\bf{A}_{k} \in \mathbb{R}^{ m \times m}$ of $\bf{A}$ as the sum of a sparse block $\bf{A}_{k}^{0}$ and the sum of $r$ rank-one contributions as follows:
\begin{equation}\label{eq:R1_Block_Jacobian}
    \bf{A}_{k} \ = \ \bf{A}_{k}^{0} \ + \ \bf{U}_{k} \bf{V}_{k}^T, \ \bf{U}_{k} \in \mathbb{R}^{m \times r}, \ \bf{V}_{k} \in \mathbb{R}^{m \times r}. 
\end{equation}
The Sherman-Morrison-Woodbury (SMW) matrix inversion formula \cite{Golubook, Hager_review} provides an analytical formula for the inverse of each block of the form (\ref{eq:R1_Block_Jacobian}). If the local matrix $\bf{C}_{k} \in \mathbb{R}^{r \times r}$ defined by:
\begin{equation*}
    \bf{C}_{k} \ := \  \bf{I}_{r \times r} \ + \ \bf{V}_{k}^T (\bf{A}_{k}^{0})^{-1} \bf{U}_{k},
\end{equation*}
where $\bf{I}_{r \times r}$ is the identity matrix in $\mathbf{R}^{r \times r}$, is invertible, then we have:
\begin{equation}\label{eq:SMW}
   \bf{A}_{k}^{-1} \ = \  (\bf{A}_{k}^{0})^{-1} \ + \ (\bf{A}_{k}^{0})^{-1} \bf{U}_{k} \bf{C}_{k}^{-1} \bf{V}_{k}^{T} (\bf{A}_{k}^{0})^{-1}.
\end{equation}
A comprehensive discussion of this result can be found in Hager \cite{Hager_review}. This formula offers an alternative strategy to solve the local system $\bf{A}_{k} \bf{x}_{k} = \bf{b_{k}}$:
\begin{enumerate}
    \item Compute the solutions $\bf{x}_{k}^{0} \in \mathbb{R}^{m}$ and $\bf{W}_{k} \in \mathbb{R}^{m \times r}$ of the sparse systems:
    \begin{equation}\label{eq:SMW_1}\tag{SMW-1}
        \bf{A}_{k}^{0} \bf{x}_{k} \ = \ \bf{b}_{k} \ \mbox{and} \  \bf{A}_{k}^{0} \bf{W}_{k} \ = \ \bf{U}_{k},
    \end{equation}
respectively.
    \item Compute the solution $\bf{z}_{k} \in \mathbb{R}^{r}$ to the reduced linear system:
    \begin{equation}\label{eq:SMW_2} \tag{SMW-2}
        \bf{C}_{k} \bf{z}_{k} \ = \ \bf{V}_{k}^{T} \bf{x}_{k}^{0}.
    \end{equation}
    \item Assemble the the final solution $\bf{x}_{k} \ := \ \bf{x}_{k}^{0} \ + \ \bf{W}_{k} \bf{z}_{k}$.
\end{enumerate}
For this strategy to be effective, the combined cost of solving $(r+1)$ sparse systems (\ref{eq:SMW_1}) and solving the reduced system (\ref{eq:SMW_2}) must be faster than solving the dense $m \times m$ system $\bf{A}_{k} \bf{x}_{k} = \bf{b}_{k}$ directly.

%
%
%

\section{Conclusions}
\myindent In this work, we introduced an alternative data structure to store and operate on Jacobian matrices arising in implicit simulations of coupled inviscid fluid and chemical processes. This data structure is motivated by the challenges posed by the simulation of entry physics phenomena using advanced CFD algorithms that integrate higher-fidelity thermo-chemistry models such as State-to-State kinetics. These models dramatically increase the number of governing equations because species are now tracked according to quantum energy levels. Both the temporal and the spatial complexity of Jacobian operations grows quadratically with respect to the number of species. The same complexities become linear with our data structure. \\
\myindent The key idea of this work is to recognize that for such systems, the block-dense structure of the Jacobian is in large part due to a number of rank-one contributions that is relatively small compared to the total number of equations. Linear algebra operations on rank-one matrices can easily be done in linear time. This leads to an interesting configuration where using two sparse data structures to represent a Jacobian block is better than using a dense matrix. Contributions from inviscid interior trace residual terms meet these conditions for HLL-type fluxes. A similar breakdown arises when considering transition processes between quantum levels. Contributions to temporal source Jacobian entries caused by pure collision partners can be grouped into rank-one contributions. \\
\myindent The present work is limited in scope in that first-order implicit schemes are considered, and that the underlying physics must have a certain structure. The use of a block r1-sparse Jacobian representation might not be possible for physical models integrating non-sparse dynamics such as multicomponent diffusion \cite{Cook, Giovangigli}. Regarding higher-order discretizations, variations of the block r1-sparse representation could be possible depending on how high-order accuracy is achieved by the scheme. For high-order Discontinuous Galerkin (DG) schemes for example, efficient preconditioners have been developed by leveraging tensor-product bases \cite{Diosady}. An extension of the block r1-sparse representation could make use of similar ideas. \\
\myindent Representing the Jacobian of a CFD discretization in the proposed way might have benefits beyond our specific application. The r1-sparse representation is valid irrespective of the number of species in the system. Furthermore, it expresses a richer structure by separating, within each block, contributions according to their locality. This richness could be leveraged into smarter preconditioning techniques, such as the variation of the block-Jacobi technique using the SMW matrix identity \cite{Hager_review} discussed in section 7.

\section*{Acknowledgments}
\myindent This research was funded by a NASA Postdoctoral Fellowship administered through Oak Ridge Associated Universities (ORAU). Support from the NASA Space Technology Mission Directorate (STMD) through the Entry Systems Modeling (ESM) project is gratefully acknowledged. \\
\myindent Ayoub Gouasmi is grateful to John Biddiscombe for valuable discussions and lectures about generic programming during a C++ training event hosted by the Swiss Supercomputing Center in October 2023. Ayoub Gouasmi is also grateful to Hilario Torres and Levi Barnes for productive discussions during the 2023 NASA GPU Hackathon.

\end{document}